\documentclass[adp,a4paper,fleqn%
]{w-art}
\usepackage{times,cite,w-thm}

\def\beqa{\begin{eqnarray}}
\def\eeqa{\end{eqnarray}}
\def\beq{\begin{equation}}
\def\eeq{\end{equation}}

\def\de {{\delta}}

\def \phan{\phantom}

\def\ie{{\it i.e. }}

\def\pl{{\it Phys. Lett.}\ }

\def\rmp{{\it Rev. Mod. Phys.}\ }

\catcode`\@=12

\newtheorem{Proposition}{Proposition}[section]

\renewcommand{\theenumi}{\alph{enumi}}
\renewcommand{\theenumii}{\roman{enumi}}

\newfont{\gotico}{eufm10 scaled\magstephalf}
\newfont{\qvd}{msam10 scaled\magstephalf}

\def\de#1/de#2{\frac{\partial {#1}}{\partial {#2}}}
\def\De#1/de#2{\dfrac{\partial {#1}}{\partial {#2}}}

\def\const{{\rm const.}}

\theoremstyle{plain}

\theoremstyle{definition}

\usepackage[]{graphicx}
\begin{document}

\DOIsuffix{theDOIsuffix}

\Volume{1} \Month{01} \Year{2007}

\pagespan{1}{}

\Receiveddate{XXXX} \Reviseddate{XXXX} \Accepteddate{XXXX}
\Dateposted{XXXX}



\title[$f(R)$-gravity with torsion]{Metric-affine $f(R)$-gravity with torsion: an overview}

\author[S. Capozziello]{Salvatore Capozziello\inst{1}
  \footnote{Corresponding author\quad
  E-mail:~\textsf{capozziello@na.infn.it}}}
\address[\inst{1}]{Dipartimento di Scienze Fisiche\ Universit\`a ``Federico II'' di Napoli and INFN Sez. di Napoli
        Compl. Univ. Monte S. Angelo Ed. N, via Cinthia I- 80126 Napoli (Italy)}

\author[S. Vignolo]{Stefano Vignolo\inst{2}}
\address[\inst{2}]{DIPTEM Sez. Metodi e Modelli Matematici, Universit\`a di Genova
                Piazzale Kennedy, Pad. D - 16129 Genova (Italy)}

\begin{abstract}
Torsion and curvature could play a fundamental role in explaining
cosmological  dynamics.  $f(R)$-gravity with torsion is an
approach aimed to encompass in a comprehensive scheme all the Dark
Side of the Universe (Dark Energy and Dark Matter). We discuss the
field equations in empty space and in presence of perfect fluid
matter taking into account the analogy with the metric-affine
formalism. The result is that the extra curvature and torsion
degrees of freedom can be dealt under the standard of an effective
scalar field of fully geometric origin. The initial value problem
for such theories is also discussed.
\end{abstract}

\maketitle

\section{Introduction}
The issue to enlarge or revise  the General Relativity (GR) comes
out since several questions in modern physics could strictly
depend on the fact that it is a classical theory. This means that
quantum effects (ultra-violet scales) are not considered in GR.
Besides, it seems to fail at galactic and cosmological scales
(infra-red scales) leading to the disturbing puzzles of Dark
Energy and Dark Matter. Quantum effects should be considered in
any theory which deals with gravity at a fundamental level.
Assuming a {\bf U}$_4$ manifold, instead of {\bf V}$_4$,  is the
first straightforward generalization which tries to include the
spin fields into the  geometrical framework of GR. The
Einstein--Cartan--Sciama--Kibble (ECSK) theory is one of the most
serious attempts in this direction \cite{hehl}. However, this mere
inclusion of spin matter fields does not exhaust the role of
torsion which  gives important contributions in any fundamental
theory. For example, a torsion field appears in (super)string
theory, if we consider the fundamental string modes; we need, at
least, a scalar mode and two tensor modes: a symmetric and an
antisymmetric one. The latter one, in the low energy limit, is a
torsion field \cite{GSW}. Furthermore, several attempts of
unification between gravity and electromagnetism have to take into
account torsion in four and in higher--dimensional theories such
as Kaluza-Klein ones \cite{GER}. Any theory of gravity considering
twistors needs the inclusion of torsion \cite{HOW} while
Supergravity is the natural arena in which torsion, curvature and
matter fields are treated under the same standard  \cite{LOS}.

Besides, several people agree with the line of thinking that
curvature and torsion could  play  specific roles in cosmological
dynamics, at early and late epochs.  In fact, the interplay of
curvature and torsion naturally gives repulsive contributions to
the energy--momentum tensor so that cosmological models become
singularity--free and accelerating \cite{SIV}.

All these arguments do not allow to neglect torsion in any
comprehensive theory of gravity which takes into account the
non--gravitational counterpart of the fundamental interactions.
However, in most papers, a clear distinction is not made among the
different kinds of torsion. Usually torsion is simply related to
the spin--density of matter, but, very often,  it assumes more
general meanings. It can be shown that there are many independent
torsion tensors with different properties
\cite{Capozziello:2001mq}.

On the other hand, the issue to extended GR to more general
actions with respect to the Hilbert-Einstein one is revealing a
very fruitful approach. From a conceptual point of view, there is
no {\it a priori} reason to restrict the gravitational Lagrangian
to a linear function of the Ricci scalar $R$, minimally coupled
with matter. The idea that there are no ``exact'' laws of physics
but that the Lagrangians of physical interactions are
``stochastic'' functions -- with the property that local gauge
invariances (\ie conservation laws) are well approximated in the
low energy limit and that physical constants can vary -- has been
taken into serious consideration \cite{ottewill}.

Beside fundamental physics motivations, all these theories have
acquired a huge interest in cosmology due to the fact that they
``naturally" exhibit inflationary behaviors able to overcome the
shortcomings of Cosmological Standard Model \cite{starobinsky}.
Furthermore, dark energy models mainly rely on the implicit
assumption that Einstein's GR is the correct theory of gravity,
indeed. Nevertheless, its validity at the larger astrophysical and
cosmological scales has never been tested \cite{will}, and it is
therefore conceivable that both cosmic speed up and missing matter
represent signals of a breakdown in our understanding of
gravitation law so that one should consider the possibility that
the Hilbert\,-\,Einstein Lagrangian, linear in the Ricci scalar
$R$, should be generalized. Following this line of thinking, the
choice of a generic function $f(R)$ can be derived by matching the
data and by the "economic" requirement that no exotic ingredients
have to be added. This is the underlying philosophy of what is
referred to as $f(R)$--gravity (see \cite{reviews} and references
therein). In this context, the same cosmological constant could be
removed as an ingredient of the cosmic pie being nothing else but
a particular eigenvalue of a general class of theories
\cite{garattini}.

However $f(R)$--gravity can be encompassed in the Extended Theories
of Gravity being a "minimal" extension of GR where (analytical)
functions of Ricci scalar are taken into account. Although higher
order gravity theories have received much attention in cosmology,
since they are naturally able to give rise to  accelerating
expansions (both in the late and in the early Universe) and
systematic studies of the phase space of solutions are in progress
\cite{cnot}, it is possible to demonstrate that $f(R)$--theories
can also play a major role at astrophysical scales. In fact,
modifying the gravity Lagrangian can affect the gravitational
potential in the low energy limit. Provided that the modified
potential reduces to the Newtonian one at Solar System scales,
this implication could represent an intriguing opportunity rather
than a shortcoming for $f(R)$--theories \cite{noiaa}. A corrected
gravitational potential could offer the possibility to fit the
galaxy rotation curves and the density profiles of galaxy clusters
without the need of dark matter \cite{mnras1,mnras2}.

In this paper, we want to give an overview of the main features of
$f(R)$--gravity theories considering also torsion. After reviewing
the basic definitions of torsion and outlying the possibility of a
classification in Sec.2, we discuss the field equations of
$f(R)$--gravity with torsion {\it in vacuo} (Sec.3) and in presence
of perfect fluid matter (Sec.4). Analogies and differences with
respect to the Palatini approach in ${\bf V}_4$ are stressed. The
equivalence of such theories with Scalar--Tensor Gravity is
discussed in Sec.5, while the Cauchy problem for $f(R)$--gravity in
presence of perfect-fluid matter is discussed in Sec.6.  Conclusions
are drawn in Sec.7.

\section{ Generalities on torsion}
In this section, we give general features of torsion and
associated quantities that are defined in  {\bf U}$_{4}$
space-times \cite{hehl}. After, we will apply this formalism in
the framework of  $f(R)$--gravity.

Torsion tensor $S_{ab}^{\phan{ab}c}$ is the antisymmetric part
of the affine connection coefficients $\Gamma_{ab}^{c}$, that is
 \beq
 \label{t1}
S_{ab}^{\phan{ab}c}=\frac{1}{2}\left(\Gamma_{ab}^{c}-\Gamma_{ba}^{c}\right)
\equiv\Gamma_{[ab]}^{c}\,, \eeq where $a,b,c = 0,\dots 3$. In GR,
it is postulated that $S_{ab}^{\phan{ab}c}=0$. It is a general
convention to call {\bf U}$_4$ a $4$-dimensional space-time
manifold endowed with metric and torsion. The manifolds with
metric and without torsion are labelled as {\bf V}$_4$.  In
general, torsion occurs in linear combinations as the {\it
contortion tensor}
 \beq
\label{t3}
K_{ab}^{\phantom{ab}c}=-S_{ab}^{\phantom{ab}c}-S^{c}_{\phan{c}a b}
+
S^{\phantom{a}c}_{b\phantom{c}{a}}=-K^{\phantom{a}c}_{a\phan{c}b}\,,
\eeq and  the {\it modified torsion tensor}

\beq \label{t4} T_{ab}^{\phantom{ab}c}=
S_{ab}^{\phantom{ab}c}+2\delta^{\phan{[a}c}_{[a}S_{b]}
 \eeq
where $S_a\equiv S_{ab}^{\phantom{ab}b}$. According to these
definitions,  the affine connection can be written as \beq
\label{t5}
\Gamma_{ab}^{c}=\left\{^{c}_{ab}\right\}-K_{ab}^{\phantom{ab}c}\,,
\eeq where $\left\{^{c}_{ab}\right\}$ are the usual Christoffel
symbols of the symmetric connection. The presence of torsion in
the affine connection implies that the covariant derivatives of a
scalar field $\phi$ do not commute, that is
\begin{equation}\label{scalarder}
 \tilde \nabla_{[a}\tilde\nabla_{b]}\phi=-S_{ab}^{\phan{ab}c}\tilde\nabla_{c}\phi;
\end{equation}
and for a vector $v^a$ and a covector $w_a$,  the following
relations
\begin{equation}\label{doppiader1}
 (\tilde\nabla_{a}\tilde\nabla_{b} - \tilde\nabla_{b}\tilde\nabla_{a})v^{c}=
  R_{abd}^{\phantom{abd}c}v^d
 -2S_{ab}^{\phan{ab}d}\tilde\nabla_{d}v^c,
 \end{equation}
 and
\begin{equation}\label{doppiader2}
 (\tilde\nabla_{a}\tilde\nabla_{b} - \tilde\nabla_{b}\tilde\nabla_{a})w_{d} =
  R_{abc}^{\phantom{abc}d}w_{d}
 -2S_{ab}^{\phan{ab}d}\tilde\nabla_{d}w_{c}
\end{equation}
hold. Here $R_{abc}^{\phantom{abc}d}$ is the Riemann tensor. The
contribution to the Riemann tensor of torsion can be explicitly
given by
\begin{equation}\label{riexpanded}
R_{abc}^{\phantom{abd}d} =R_{abc}^{\phantom{abd}d}(\{\}) -
 \nabla_{a}K_{bc}^{\phantom{b]c}d} +  \nabla_{b}K_{ac}^{\phantom{ac}d}
+ K_{ae}^{\phantom{ae}d}K_{bc}^{\phantom{bc}e}-
K_{be}^{\phantom{be}d}K_{ac}^{\phantom{ac}e}
\end{equation}
where $R_{abc}^{\phantom{abc}d}(\{\})$ is the tensor generated by
the Christoffel symbols. The symbols $\tilde\nabla$ and $\nabla$
have been used to indicate the covariant derivative with and
without torsion respectively. From Eq.(\ref{riexpanded}), the
expressions for the Ricci tensor and the curvature scalar are
\begin{equation}\label{ricci}
 R_{ab}= R_{ab}(\{\}) - 2\nabla_{a}S_{c} + \nabla_{b}K_{ac}^{\phantom{ac}b}
+ K_{ae}^{\phantom{ae}b}K_{bc}^{\phantom{bc}e}-
2S_eK_{ac}^{\phantom{ac}e}
\end{equation}
and
\begin{equation}\label{curvscalar}
 R=R(\{\}) - 4 \nabla_{a}S^{a} + K_{ceb}K^{bce} - 4 S_aS^a.
\end{equation}
Torsion can be decomposed with respect to the Lorentz group into
three irreducible tensors
\begin{equation}\label{decomposition}
 S_{ab}^{\phantom{ab}c}= {}^T S_{ab}^{\phantom{ab}c} +
{}^A S_{ab}^{\phantom{ab}c}+ {}^V S_{ab}^{\phantom{ab}c}.
\end{equation}
Torsion has $24$ components, of which ${}^T S_{ab}$ has $16$
components, ${}^A S_{ab}$ has $4$ and ${}^V S_{ab}$ has the
remaining $4$ components. One has
\begin{equation}\label{vector}
 {}^V S_{ab}^{\phantom{ab}c}=\frac{1}{3}(S_{a}\delta_{b}^{c}-
 S_{b}\delta_{a}^{c}),
\end{equation}
where $S_a =  S_{ab}^{\phantom{ab}b}$,
\begin{equation}\label{axial}
 {}^A S_{ab}^{\phantom{ab}c}=g^{cd}S_{[abd]}
\end{equation}
that is called the axial (or totally anti-symmetric) torsion, and
\begin{equation}\label{tensor}
 {}^T S_{ab}^{\phantom{ab}c}= S_{ab}^{\phantom{ab}c} - {}^A
 S_{ab}^{\phantom{ab}c}- {}^V S_{ab}^{\phantom{ab}c}
\end{equation}
that is the traceless non-totally anti-symmetric part of torsion.
It is immediately clear that relating torsion to the spin-density
of matter is only one of the possible aspects of the problem
\cite{Capozziello:2001mq}.

\section{Field equations in vacuo}
In the metric-affine formulation of $f(R)$-gravity, the dynamical
fields are pairs $(g,\Gamma)\/$ where $g\/$ is the metric and
$\Gamma\/$ is the linear connection. Considering torsion means to
assume a general non-symmetric connection. In vacuo, the field
equations are obtained by varying with respect to the metric and
the connection the following action
\begin{equation}\label{1.0}
{\cal A}\/(g,\Gamma)=\int{\sqrt{|g|}f\/(R)\,ds}
\end{equation}
where $f(R)$ is a real function, $R\/(g,\Gamma) = g^{ij}R_{ij}\/$
(with $R_{ij}:= R^h_{\;\;ihj}\/$) is the scalar curvature
associated to the dynamical connection $\Gamma\/$. More precisely,
in the approach with torsion, one can ask for a metric connection
$\Gamma$ with torsion different from zero while, in the Palatini
approach,  $\Gamma$ is non-metric but torsion is null. In
vacuo, the field equations for $f(R)$-gravity with torsion are
\cite{CCSV1,CCSV3,CCSV2}
\begin{subequations}\label{1.1}
\begin{equation}\label{1.1a}
f'\/(R)R_{ij} - \frac{1}{2}f\/(R)g_{ij}=0\,,
\end{equation}
\begin{equation}\label{1.1b}
T_{ij}^{\;\;\;h} = -
\frac{1}{2f'}\de{f'}/de{x^p}\/\left(\delta^p_i\delta^h_j -
\delta^p_j\delta^h_i\right)\,,
\end{equation}
\end{subequations}
while the field equations for $f(R)$-gravity  {\it \`{a} la}
Palatini are
\begin{subequations}\label{1.2}
\begin{equation}\label{1.2a}
f'\/(R)R_{ij} - \frac{1}{2}f\/(R)g_{ij}=0\,,
\end{equation}
\begin{equation}\label{1.2b}
\nabla_k\/(f'(R)g_{ij})=0\,.
\end{equation}
\end{subequations}
In both cases, considering the trace of  Einstein-like field
equations  \eqref{1.1a} and \eqref{1.2a}, one gets the relation
\begin{equation}\label{1.3}
f'\/(R)R  - 2f\/(R)=0\,.
\end{equation}
The latter is an identity automatically satisfied by all possible
values of $R\/$ only in the special case $f\/(R)=\alpha R^2\/$. In
all other cases, equation \eqref{1.3} represents a constraint on
the scalar curvature $R\/$. Therefore, if $f\/(R)\not = \alpha
R^2\/$, the scalar curvature $R$ is a constant coinciding with the
solution of Eq.\eqref{1.3}. In this case, Eqs. \eqref{1.1b} and
\eqref{1.2b} imply that both dynamical connections coincide with
the  Levi--Civita connection  associated to the metric $g_{ij}$
that is solution of the Einstein-like field equations. In other
words, both theories reduce to the Einstein theory with
cosmological constant. On the contrary, only in the case $f\/(R) =
\alpha R^2\/$ metric-affine $f(R)\/$-theories differ from GR  in
vacuo. To see this point, let us consider  the theory with
torsion: from Eqs. \eqref{1.1},  we obtain final field equations
of the form
\begin{equation}\label{00.20}
\tilde{R}_{ij} - \frac{1}{4}\tilde{R}g_{ij} =
-\frac{2}{3}\tilde{\nabla}_jT_i +
\frac{1}{6}\tilde{\nabla}_hT^hg_{ij} -\frac{2}{9}T_iT_j +
\frac{1}{18}T_hT^hg_{ij}
\end{equation}
and
\begin{equation}\label{00.21}
\de /de{x^i}\/\left( \tilde{R} + 2\tilde{\nabla}_hT^h -
\frac{2}{3}T_hT^h \right) = -\frac{2}{3}\left( \tilde{R} +
2\tilde{\nabla}_hT^h - \frac{2}{3}T_hT^h \right)\/T_i
\end{equation}
where $\tilde{R}_{ij}\/$ and $\tilde{R}\/$ are respectively the
Ricci tensor and scalar associated with the metric $g_{ij}\/$,
$\tilde{\nabla}_h\/$ denotes the Levi--Civita covariant derivative
induced by $g_{ij}\/$ and $T_i :=T_{ij}^{\;\;j}$ is the trace of the above defined modified torsion tensor.

\section{Field equations in presence of matter}
In presence of matter, the field equations are derived from an
action functional of the form
\begin{equation}\label{2.1}
{\cal A}\/(g,\Gamma)=\int{\left(\sqrt{|g|}f\/(R) + {\cal
L}_m\right)\,ds}
\end{equation}
where ${\cal L}_m\\/$ is a suitable material Lagrangian density.
Assuming that the material Lagrangian does not depend on the
dynamical connection, the field equations are
\begin{subequations}\label{2.2}
\begin{equation}\label{2.2a}
f'\/(R)R_{ij} - \frac{1}{2}f\/(R)g_{ij}=\Sigma_{ij}\,,
\end{equation}
\begin{equation}\label{2.2b}
T_{ij}^{\;\;\;h} = -
\frac{1}{2f'\/(R)}\de{f'\/(R)}/de{x^p}\/\left(\delta^p_i\delta^h_j
- \delta^p_j\delta^h_i\right)
\end{equation}
\end{subequations}
for $f(R)$-gravity with torsion, and
\begin{subequations}\label{2.3}
\begin{equation}\label{2.3a}
f'\/(R)R_{ij} - \frac{1}{2}f\/(R)g_{ij}=\Sigma_{ij}\,,
\end{equation}
\begin{equation}\label{2.3b}
\nabla_k\/(f'(R)g_{ij})=0\,,
\end{equation}
\end{subequations}
for $f(R)$-gravity in the  Palatini approach. In Eqs. \eqref{2.2a}
and \eqref{2.3a}, the quantity ${\displaystyle \Sigma_{ij}:= -
\frac{1}{\sqrt{|g|}}\frac{\delta{\cal L}_m}{\delta g^{ij}}\/}$ is
the stress-energy tensor. Considering the trace of Eqs.
\eqref{2.2a} and \eqref{2.3a}, we obtain a relation between the
curvature scalar $R\/$ and the trace of the stress-energy tensor
$\Sigma:=g^{ij}\Sigma_{ij}\/$. We have indeed
\begin{equation}\label{2.4}
f'\/(R)R -2f\/(R) = \Sigma\,.
\end{equation}
From now on, we shall suppose that the relation \eqref{2.4} is
invertible as well as that $\Sigma\not=\const\/$ Under these
hypotheses, the curvature scalar $R\/$ can be expressed as a
suitable function of $\Sigma\/$, namely
\begin{equation}\label{2.5}
R=F(\Sigma)\,.
\end{equation}
Starting from this result and defining the scalar field
\begin{equation}\label{2.6}
\varphi:=f'(F(\Sigma))\,,
\end{equation}
we can put the field equations of both Palatini and with torsion
theories in the same form \cite{CCSV1,Olmo}, that is
\begin{equation}\label{2.7}
\begin{split}
\tilde{R}_{ij} -\frac{1}{2}\tilde{R}g_{ij}=
\frac{1}{\varphi}\Sigma_{ij} + \frac{1}{\varphi^2}\left( -
\frac{3}{2}\de\varphi/de{x^i}\de\varphi/de{x^j}
+ \varphi\tilde{\nabla}_{j}\de\varphi/de{x^i} + \frac{3}{4}\de\varphi/de{x^h}\de\varphi/de{x^k}g^{hk}g_{ij} \right. \\
\left. - \varphi\tilde{\nabla}^h\de\varphi/de{x^h}g_{ij} -
V\/(\varphi)g_{ij} \right)\,,
\end{split}
\end{equation}
where we have introduced the effective potential
\bigskip\noindent
\begin{equation}\label{2.8}
V\/(\varphi):= \frac{1}{4}\left[ \varphi
F^{-1}\/((f')^{-1}\/(\varphi)) +
\varphi^2\/(f')^{-1}\/(\varphi)\right]\,,
\end{equation}
for the scalar field $\varphi\/$. In Eq. \eqref{2.7},
$\tilde{R}_{ij}\/$, $\tilde{R}\/$ and $\tilde\nabla\/$ denote
respectively the Ricci tensor, the scalar curvature and the
covariant derivative associated with the Levi--Civita connection
of the dynamical metric $g_{ij}$. Therefore, if the dynamical
connection $\Gamma\/$ is not coupled with matter, both 
theories (with torsion and Palatini--like) generate identical
Einstein--like field equations. On the contrary, the field
equations for the dynamical connection are different and (in
general) give rise to different solutions. In fact, the connection
$\Gamma$ solution of Eqs. \eqref{2.2b} is
\begin{equation}\label{2.9}
\Gamma_{ij}^{\;\;\;h} =\tilde{\Gamma}_{ij}^{\;\;\;h} +
\frac{1}{2\varphi}\de\varphi/de{x^j}\delta^h_i -
\frac{1}{2\varphi}\de\varphi/de{x^p}g^{ph}g_{ij}
\end{equation}
where $\tilde{\Gamma}_{ij}^{\;\;\;h}\/$ denote the coefficients of
the Levi--Civita connection associated with the metric $g_{ij}$,
while the connection $\bar\Gamma\/$ solution of Eqs. \eqref{2.3b}
is
\begin{equation}\label{2.10}
\bar{\Gamma}_{ij}^{\;\;\;h}= \tilde{\Gamma}_{ij}^{\;\;\;h} +
\frac{1}{2\varphi}\de\varphi/de{x^j}\delta^h_i -
\frac{1}{2\varphi}\de\varphi/de{x^p}g^{ph}g_{ij} +
\frac{1}{2\varphi}\de\varphi/de{x^i}\delta^h_j\,,
\end{equation}
and coincides with the Levi--Civita connection induced by the
conformal metric $\bar{g}_{ij}:=\varphi g_{ij}$. By comparison,
the connections $\Gamma$ and $\bar\Gamma$ satisfy the relation
\begin{equation}\label{2.11}
\bar{\Gamma}_{ij}^{\;\;\;h} =\Gamma_{ij}^{\;\;\;h} +
\frac{1}{2\varphi}\de\varphi/de{x^i}\delta^h_j\,.
\end{equation}
Of course, the Einstein--like equations \eqref{2.7} are coupled
with the matter field equations. In this respect, it is worth
pointing out that Eqs. \eqref{2.7} imply the same conservation
laws holding in GR. We have in fact \cite{CV1,CV2}
\begin{Proposition}\label{ProB.1}
Eqs. \eqref{2.6}, \eqref{2.7} and \eqref{2.8} imply the standard
conservation laws $\tilde{\nabla}^j\/\Sigma_{ij}=0$.
\end{Proposition}
Summarizing all the  results, we can say that if both the theories
(with torsion and Palatini--like) are considered as ``metric'', in
the sense that the dynamical connection $\Gamma\/$ is not coupled
with matter ${\displaystyle \left(\frac{\delta{\cal
L}_m}{\delta\Gamma}=0\right)}$ and it does not define parallel
transport and covariant derivative in space--time, then the two
approaches are completely equivalent. Indeed, in the ``metric''
framework, the true connection of space--time is the Levi--Civita
one associated with the metric $g\/$ and the role played by the
dynamical connection $\Gamma\/$ is just to generate the right
Einstein--like equations of the theory. On the contrary, if the
theories are genuinely metric--affine, then they are different
even though the condition ${\displaystyle \frac{\delta{\cal
L}_m}{\delta\Gamma}=0\/}$ is satisfied.

At this point, some considerations are due in relation to conformal transformations.
If the trace $\Sigma$ of the stress-energy tensor is independent
of the metric $g_{ij}$, Eqs.\eqref{2.7} can be
simplified by passing from the Jordan to the Einstein frame.
Indeed, performing the conformal transformation
$\bar{g}_{ij}=\varphi\/g_{ij}$, it is easily seen that Eqs.
\eqref{2.7} assume the simpler form (see also
\cite{CCSV1,Olmo})
\begin{equation}\label{3.1}
\bar{R}_{ij} - \frac{1}{2}\bar{R}\bar{g}_{ij} =
\frac{1}{\varphi}\Sigma_{ij} -
\frac{1}{\varphi^3}V\/(\varphi)\bar{g}_{ij}\,,
\end{equation}
where $\bar{R}_{ij}\/$ and $\bar{R}\/$ are respectively the Ricci
tensor and the curvature scalar derived from the conformal metric
$\bar{g}_{ij}\/$. As mentioned above, the conformal transformation
is working if the trace of the stress-energy tensor is independent
of the metric $g_{ij}$. Only in this case, Eqs.
\eqref{3.1} depend exclusively on the conformal metric
$\bar{g}_{ij}$ and the other matter fields. The quantity
\begin{equation}\label{3.2}
T_{ij}:=\frac{1}{\varphi}\Sigma_{ij} - \frac{1}{\varphi^3}V\/(\varphi)\bar{g}_{ij},
\end{equation}
appearing in the right--hand side of Eqs. \eqref{3.1},  plays the
role of effective stress--energy tensor for the conformally
transformed theory. The existing relation between the conservation
laws, holding in the Jordan and in the Einstein frame separately,
is explained by the following \cite{CV1,CV2}
\begin{Proposition}\label{ProB.2}
Given the Levi-Civita connection $\bar\Gamma\/$
derived from the conformal metric tensor $\bar{g}\/$
and given the associated covariant derivative $\bar\nabla\/$,
the condition $\bar{\nabla}^j\/T_{ij}=0\/$ is equivalent to the
condition $\tilde{\nabla}^j\/\Sigma_{ij}=0$.
\end{Proposition}
This result will be particularly useful for the well--formulation and
 the well--position of the Cauchy problem.

\section{Equivalence with scalar--tensor theories}
The action functional of a (purely metric) scalar--tensor theory
is given by
\begin{equation}\label{00001}
{\cal A}\/(g,\varphi)=\int{\left[\sqrt{|g|}\left(\varphi\tilde{R}
-\frac{\omega_0}{\varphi}\varphi_i\varphi^i - U\/(\varphi)
\right)+ {\cal L}_m\right]\,ds}
\end{equation}
where $\varphi\/$ is the scalar field, $\varphi_i :=
\de\varphi/de{x^i}\/$ and $U\/(\varphi)\/$ is the potential of
$\varphi\/$. The matter Lagrangian ${\cal L}_m\/(g_{ij},\psi)\/$
is a function of the metric and some matter fields $\psi\/$.
$\omega_0\/$ is the so called Brans--Dicke parameter. The field
equations, derived by varying with respect to the metric and the
scalar field, are
\begin{equation}\label{0000.2}
\tilde{R}_{ij} -\frac{1}{2}\tilde{R}g_{ij}=
\frac{1}{\varphi}\Sigma_{ij} + \frac{\omega_0}{\varphi^2}\left(
\varphi_i\varphi_j  - \frac{1}{2}\varphi_h\varphi^h\/g_{ij}
\right) + \frac{1}{\varphi}\left( \tilde{\nabla}_{j}\varphi_i -
\tilde{\nabla}_h\varphi^h\/g_{ij} \right) -
\frac{U}{2\varphi}g_{ij}
\end{equation}
and
\begin{equation}\label{0000.3}
\frac{2\omega_0}{\varphi}\tilde{\nabla}_h\varphi^h + \tilde{R} -
\frac{\omega_0}{\varphi^2}\varphi_h\varphi^h - U' =0
\end{equation}
where $\Sigma_{ij}:= - \frac{1}{\sqrt{|g|}}\frac{\delta{\cal
L}_m}{\delta g^{ij}}\/$ and $U' :=\frac{dU}{d\varphi}\/$. Taking
the trace of Eq. \eqref{0000.2} and using it to replace $\tilde
R\/$ in Eq. \eqref{0000.3}, one obtains the equation
\begin{equation}\label{0000.4}
\left( 2\omega_0 + 3 \right)\/\tilde{\nabla}_h\varphi^h = \Sigma +
\varphi U' -2U\,.
\end{equation}
By a direct comparison, it is immediately seen that for
\begin{equation}
\omega_0 =-\frac{3}{2} \quad {\rm and} \quad U\/(\varphi) =\frac{2}{\varphi}V\/(\varphi) 
\end{equation}
Eqs. \eqref{0000.2} become formally identical to the
Einstein--like equations \eqref{2.7} for a metric--affine $f(R)\/$
theory. Moreover, in such a circumstance, Eq. \eqref{0000.4}
reduces to the algebraic equation
\begin{equation}\label{0000.5}
\Sigma + \varphi U' -2U =0\,,
\end{equation}
relating the matter trace $\Sigma\/$ to the scalar field
$\varphi\/$, just as it happens for $f(R)\/$ theories. More in
particular, it is a straightforward matter to verify that (under
the condition $f''\not= 0\/$ \cite{CCSV1}) Eq. \eqref{0000.5}
expresses exactly the inverse relation of
\begin{equation}\label{0000.6}
\varphi=f'\/(F(\Sigma)) \quad {\rm namely} \quad
\Sigma=F^{-1}\/((f')^{-1}\/(\varphi))\,,
\end{equation}
being $F^{-1}\/(X) = f'\/(X)X - 2f\/(X)\/$. In conclusion, it follows that, in the ``metric'' interpretation,
metric--affine $f(R)\/$-- theories (with torsion or Palatini) are
equivalent to scalar--tensor theories with Brans--Dicke parameter
$\omega_0 =-\frac{3}{2}\/$.
It is also possible to show the equivalence
between $f(R)\/$ theories and $\omega_0 =0\/$ scalar--tensor theories with torsion. To this end, let us consider an action functional of the kind
\begin{equation}\label{00007}
{\cal A}\/(g,\Gamma,\varphi)=\int{\left[\sqrt{|g|}\left(\varphi{R} - U\/(\varphi) \right)+ {\cal L}_m\right]\,ds}\,,
\end{equation}
where the dynamical fields are respectively a metric $g_{ij}\/$,
 a metric connection $\Gamma_{ij}^{\;\;\;k}\/$ and a 
scalar field $\varphi\/$. Action \eqref{00007} describes a
scalar--tensor theory with torsion and parameter $\omega_0 =0\/$.
The variation with respect to $\varphi\/$ yields the first field
equation
\begin{equation}\label{00008}
R = U'\/(\varphi)\,,
\end{equation}
while variations with respect
to the metric and the connection give rise to the resulting field equations
\begin{equation}\label{0000.9}
T_{ij}^{\;\;\;h} = - \frac{1}{2\varphi}\de{\varphi}/de{x^p}\/\left(\delta^p_i\delta^h_j - \delta^p_j\delta^h_i\right)
\end{equation}
and
\begin{equation}\label{0000.10}
R_{ij} -\frac{1}{2}Rg_{ij}= \frac{1}{\varphi}\Sigma_{ij} - \frac{1}{2\varphi}U\/(\varphi)g_{ij}\,.
\end{equation}
Inserting the content of Eq. \eqref{00008} in the trace of Eq. \eqref{0000.10}, we obtain again the algebraic ralation between $\Sigma\/$ and $\varphi\/$, that is
\begin{equation}\label{0000.11}
\Sigma - 2U\/(\varphi) + \varphi\/U'\/(\varphi) =0
\end{equation}
identical to Eq. \eqref{0000.5}. As above, choosing the potential
$U\/(\varphi) =\frac{2}{\varphi}V\/(\varphi)\/$, it is easily seen
that Eq. \eqref{0000.11} is equivalent to the relation
$\varphi=f'\/(F(\Sigma))\/$ defining the  scalar field $\varphi$
in terms of the trace $\Sigma$. In view of this, decomposing
$R_{ij}\/$ and $R\/$ in their Christoffel and torsion dependent
terms, Eqs. \eqref{0000.9} and \eqref{0000.10} become identical to
the field equations  \eqref{2.2}. As mentioned previously,
this fact shows the equivalence between $f(R)\/$ theories and
$\omega_0 =0\/$ scalar--tensor theories with torsion, in the
metric--affine framework.

\section{The Cauchy problem in presence of perfect fluid matter}
Let us consider now metric--affine $f(R)\/$ theories coupled with
a perfect fluid. We shall show that, in the Einstein frame, the
analysis of the Cauchy problem can be carried out  following the
same arguments developed in \cite{yvonne}. The result is
that metric--affine $f(R)\/$ theories, coupled with perfect fluid
matter, possess a well formulated and well posed Cauchy
 problem depending on the form of $f(R)$.

To see this point, we start by looking for a metric $g_{ij}\/$ of signature $(-+++)\/$ in the Jordan frame, solution of Eqs. \eqref{2.7}. The stress--energy tensor will be of the form
\begin{subequations}\label{4.1}
\begin{equation}\label{4.1a}
\Sigma_{ij}=(\rho + p)\,U_iU_j + p\,g_{ij}\,,
\end{equation}
with corresponding matter field equations given by
\begin{equation}\label{4.1b}
\tilde\nabla_j\Sigma^{ij}=0\,.
\end{equation}
\end{subequations}
In Eqs. \eqref{4.1}, the scalars $\rho$ and $p$  denote
respectively the matter--energy density and the pressure of the
fluid, while $U_i$ indicate the four velocity of the fluid,
satisfying the obvious condition $g^{ij}U_iU_j =-1\/$.
After performing the conformal transformation $\bar{g}_{ij}=\varphi g_{ij}\/$ (which is  
working since the trace $\Sigma\/$ is independet of the metric $g_{ij}\/$), we can express the field equations in the Einstein frame as
\begin{subequations}\label{4.2}
\begin{equation}\label{4.2a}
\bar{R}_{ij} - \frac{1}{2}\bar{R}\bar{g}_{ij} = T_{ij}\,,
\end{equation}
and
\begin{equation}\label{4.2b}
\bar{\nabla}_j T^{ij}=0\,,
\end{equation}
\end{subequations}
where
\begin{equation}\label{4.3}
T_{ij}=\frac{1}{\varphi}(\rho + p)\,U_iU_j + \left( \frac{p}{\varphi^2} - \frac{V(\varphi)}{\varphi^3} \right)\,\bar{g}_{ij}\,,
\end{equation}
is the effective stress--energy tensor. In view of Proposition
\ref{ProB.2}, Eqs. \eqref{4.2b} are equivalent to Eqs.
\eqref{4.1b}. This is a key point in our discussion, allowing us
to apply, to the present case, the results achieved in
\cite{yvonne}. Moreover, for simplicity, we shall suppose that the
scalar field $\varphi$ is positive, that is $\varphi > 0\/$. The
opposite case $\varphi <0\/$, differing from the former only for
some technical aspects, will be briefly discussed after. Under the
above assumption, the four velocity of the fluid in the Einstein
frame can be expressed as $\bar{U}_i = \sqrt{\varphi}U_i\/$. In
view of this, the stress--energy tensor \eqref{4.3} can be
rewritten, in terms of the four velocity $\bar{U}_i\/$, as
\begin{equation}\label{4.4}
T_{ij}=\frac{1}{\varphi^2}(\rho + p)\,\bar{U}_i\bar{U}_j + \left( \frac{p}{\varphi^2} - \frac{V(\varphi)}{\varphi^3} \right)\,\bar{g}_{ij}\,.
\end{equation}
Furthermore, introducing the effective mass--energy density
\begin{subequations}\label{4.5}
\begin{equation}\label{4.5a}
\bar{\rho}:= \frac{\rho}{\varphi^2} + \frac{V(\varphi)}{\varphi^3}
\end{equation}
and the effective pressure
\begin{equation}\label{4.5b}
\bar{p}:= \frac{p}{\varphi^2} - \frac{V(\varphi)}{\varphi^3}
\end{equation}
\end{subequations}
the stress--energy tensor \eqref{4.4} assumes the final standard form
\begin{equation}\label{4.6}
T_{ij}=\left( \bar\rho + \bar{p} \right)\,\bar{U}_i\bar{U}_j + \bar{p}\,\bar{g}_{ij}\,.
\end{equation}
It is worth noticing that, starting from an equation  of state of
the form $\rho=\rho(p)\/$ and assuming that the relation
\eqref{4.5b} is invertible $(p=p(\bar{p}))$, by composition with
Eq. \eqref{4.5a} , we derive an effective equation of state
$\bar{\rho}=\bar{\rho}(\bar{p})\/$. In addition to this, we recall
that the explicit expression of the scalar field $\varphi\/$ as
well as of the potential $V(\varphi)\/$ are directly related with
the particular form of the function $f(R)\/$. Then, the
requirement of invertibility of the relation \eqref{4.5b} together
with the condition $\varphi >0\/$ (or, equivalently, $\varphi
<0\/$) become criteria for the viability of the functions
$f(R)\/$. In other words, they provide us with precise rules of
selection for the admissible functions $f(R)\/$.
From now on, the treatment of the Cauchy problem can proceed step
by step as in \cite{yvonne}.  We only recall the conclusion stated in \cite{yvonne}, where it is
proved that the Cauchy problem for the system of differential
equations \eqref{4.2}, with stress--energy tensor given by Eq.
\eqref{4.6} and equation of state
$\bar{\rho}=\bar{\rho}(\bar{p})\/$, is well--posed if the
condition
\begin{equation}\label{4.7}
\frac{d\bar{\rho}}{d\bar{p}}\geq 1\,,
\end{equation}
is satisfied. We stress that, in order to satisfy  the condition
\eqref{4.7}, we do not need to invert explicitly the relation
\eqref{4.5b}, but more simply, we have to verify
\begin{equation}\label{4.8}
\frac{d\bar{\rho}}{d\bar{p}}=\frac{d\bar{\rho}/dp}{d\bar{p}/dp}\geq 1
\end{equation}
directly from the expressions \eqref{4.5} and the  equation of
state $\rho=\rho(p)\/$. Once again, the condition \eqref{4.8},
depending on the peculiar expressions of $\varphi$ and
$V(\varphi)\/$, is strictly related to the particular form of the
function $f(R)\/$. Then, condition \eqref{4.8} represents a
further criterion for the viability of the functions $f(R)\/$.
For the sake of completeness, we outline the case $\varphi <0\/$.  We still
suppose that the signature of the metric in the Jordan frame is
$(-+++)\/$. Therefore, the signature of the conformal metric will
be $(+---)\/$ and the four velocity of the fluid in the Einstein
frame will be $\bar{U}_i = \sqrt{-\varphi}U_i\/$.
The effective stress--energy tensor will be given now by
\begin{equation}\label{4.9}
T_{ij}=-\frac{1}{\varphi^2}(\rho + p)\,\bar{U}_i\bar{U}_j + \left( \frac{p}{\varphi^2} - \frac{V(\varphi)}{\varphi^3} \right)\,\bar{g}_{ij}= \left( \bar\rho + \bar{p} \right)\,\bar{U}_i\bar{U}_j - \bar{p}\,\bar{g}_{ij}
\end{equation}
where we have introduced the quantities
\begin{subequations}\label{4.10}
\begin{equation}\label{4.10a}
\bar{\rho}:= -\frac{\rho}{\varphi^2} - \frac{V(\varphi)}{\varphi^3}
\end{equation}
and
\begin{equation}\label{4.10b}
\bar{p}:= -\frac{p}{\varphi^2} + \frac{V(\varphi)}{\varphi^3}
\end{equation}
\end{subequations}
representing, as above,  the effective mass--energy density and  pressure. 
At this point, demonstration proceeds again
as in \cite{yvonne}, except for a technical aspect. The fact  quantity $r:=\bar{\rho}+\bar{p}=-\frac{\rho +
p}{\varphi^2}\/$ this time is negative (if, as usual, $\rho\/$ and
$p\/$ are assumed positive). Therefore, instead of using the
function $\log(f^{-2}r)$ as in \cite{yvonne}, we need to use
$\log(-f^{-2}r)$. The reader can easily verify that, with this
choice, the  arguments in \cite{yvonne} apply equally well.

\section{Discussion and Conclusions}

$f(R)$--gravity seems a viable approach to solve some shortcomings
coming from GR, in particular problems related to quantization on
curved spacetime and cosmological issues related to early Universe
(inflation) and late time dark components. It is worth noticing that the scheme of
GR is fully preserved and $f(R)$ can be considered a
straightforward extension where the gravitational action has not
to be necessarily linear in the Ricci scalar $R$.  We have
discussed the possibility that also the torsion field could play
an important role in the dynamics being the ${\bf U}_4$ manifolds
a generalization of the pseudo--Riemannian
manifolds ${\bf V}_4$ (torsionless) usually adopted in GR.  
Torsion field, in the metric-affine formalism,
plays  a fundamental role in clarifying the relations between the
Palatini and the metric approaches: it gives further degrees of
freedom which  contribute, together with curvature degrees of
freedom, to the dynamics. The aim is to achieve a self-consistent
theory where unknown ingredients as dark energy and dark matter
(up to now not detected at a fundamental level) could be
completely "geometrized". Torsion field assumes a relevant role in
presence of standard matter since it allows to establish a
definite equivalence between scalar-tensor theories and 
$f(R)$--gravity, also in relation to conformal transformations. In this
case, the Chauchy problem results well formulated and well posed
depending on the form of $f(R)$ \cite{CV2}. Furthermore, from a cosmological viewpoint, torsion
field could dynamically trigger the amount of dark components
giving a straightforward explanation of the coincidence problem
\cite{CCSV1}.

\begin{acknowledgement}
We  acknowledge our friends R. Cianci and C. Stornaiolo for useful
comments and results achieved together.
\end{acknowledgement}

\end{document}